\documentclass[12pt, a4paper]{article}
\usepackage{amsmath}
\usepackage[english]{babel}
\usepackage{amsfonts}
\usepackage{enumerate}
\usepackage[pagewise,running]{lineno}
\usepackage{placeins}
\usepackage{longtable}
\usepackage{float}
\usepackage{comment}
\usepackage{xcolor}
\usepackage{lscape}
\usepackage{graphicx}
\usepackage[utf8]{inputenc}
\usepackage{amsthm,amssymb}
\usepackage{mathrsfs}
\usepackage[top=1in, bottom=1in, left=1in, right=1in]{geometry}
\numberwithin{figure}{section}
\newtheorem{theorem}{Theorem}[section]
\newtheorem{lemma}[theorem]{Lemma}
\newtheorem{corollary}[theorem]{Corollary}
\newtheorem{proposition}[theorem]{Proposition}
\newtheorem{example}[theorem]{Example}

\newtheorem{definition}[theorem]{Definition}
\newtheorem{note}[theorem]{Note}

\DeclareMathOperator{\lcm}{lcm}

\DeclareMathOperator{\hull}{Hull{(\mathcal C)}}

\begin{document}

\title{  Euclidean Hulls of Double Cyclic Codes of Arbitrary Block Lengths
\footnotetext{Email:  shuklaawadhesh@bhu.ac.in 
 (A. K. Shukla), sachiniitk93@gmail.com (Sachin Pathak), opandey1302@bhu.ac.in 
 (O. P. Pandey), mishravipul10@bhu.ac.in (V. Mishra), upadhyay@bhu.ac.in (A. K.  Upadhyay).\\
 Corresponding author- Awadhesh Kumar Shukla.}}
	\author{  {Awadhesh Kumar Shukla}$^{1}$,  {Om Prakash Pandey}$^{1}$, {Vipul Mishra }$^{1}$, \\ {Sachin Pathak}$^{2}$, {Ashish Kumar Upadhyay}$^{1}$}
\date{
		\small{
	${}^{1}${Department of Mathematics, Banaras Hindu University, Varanasi, Uttar Pradesh 221005, \\India.}\\ ${}^{2}${Department of Mathematics, University of Allahabad, Prayagraj, Uttar Pradesh, 211002 India.}}\\
\today}
\maketitle
\begin{abstract}
 A $\mathbb{Z}_2$-double cyclic code of block length $(r,s)$ can be viewed as a $\mathbb{Z}_2[x]$-submodule of $\mathbb{Z}_2[x]/\langle x^r - 1 \rangle \times \mathbb{Z}_2[x]/\langle x^s - 1 \rangle$. In 2023, Gao et al. \cite{GWF23} investigated the Euclidean hulls of double cyclic codes over $\mathbb{Z}_2$ for both separable and non-separable cases. In the non-separable setting, they determined the hull generators under the assumption that the block lengths are coprime (i.e., $\gcd(r,s)=1$) and left the non-coprime case ($\gcd(r,s) > 1$) as an open problem. In this paper, we resolve this open problem by deriving the explicit hull generators for arbitrary block lengths $(r,s)$, recovering the coprime setting as a special case of our framework. Furthermore, we establish the necessary and sufficient conditions for non-separable double cyclic codes to be linear complementary dual (LCD) and provide several numerical examples to validate our theoretical findings.   
\end{abstract}

\section{Introduction}
The concept of the hull defined algebraically as the intersection of a linear code with its dual was initially introduced in \cite{AK90} as a tool to classify finite projective planes. Since its inception, the structural properties of code hulls have been extensively explored due to their practical implications. Importantly, the size  hull determines how fast algorithms can check if two codes are equivalent or find a code's symmetries \cite{L82, L91, L97,S97, S297}. Recently, hulls have also become very useful in quantum computing, especially for building highly effective entanglement-assisted quantum error-correcting codes (EAQECCs) \cite{PPA23,AAHM22, FFLZ20, LL20}. Motivated by these diverse applications, computing the hull dimensions of linear codes over finite fields has become a prominent research direction. In 1997, Sendrier \cite {S97} successfully enumerated the number of distinct linear codes sharing a specific hull dimension. Subsequently, Skersys \cite{S03} established a mathematical expression for the average hull dimension of cyclic codes. More recently, Sangwisut et al. \cite{SJLU15} calculated the exact Euclidean hull dimensions for both cyclic and negacyclic codes over $\mathbb{F}_q$. They also provided explicit counting formulas for cyclic codes with fixed hull dimensions, alongside further theoretical investigations into the average hull dimensions of negacyclic codes.\par
Recently, Gao et al.~\cite{GWF23} began the study of Euclidean hulls for double cyclic codes over $\mathbb{Z}_2$. While they completely solved the problem for separable codes, they only looked at non-separable codes with coprime odd block lengths. Finding the hulls for any arbitrary block lengths was left as an unsolved problem. Their work inspired others to explore different algebraic structures: Shukla et al. studied hulls  of $\mathbb{Z}_p\mathbb{Z}_p[v]$-cyclic codes ~\cite{SPM26} and $\mathbb{Z}_2\mathbb{Z}_4$-additive cyclic codes~\cite{SPPVU26}, Gao et al.~\cite{LHMM25} looked at hulls of  $\mathbb{Z}_4$-double cyclic codes, and Indiver et al.~\cite{DP25} worked on  hulls of separable codes over $\mathbb{Z}_{p^2}$. However, a crucial gap remains in all these studies: they did not investigate non-separable codes of arbitrary lengths. Motivated by this missing piece, our paper directly solves this open problem. Our main goal is to find the exact generator polynomials for the Euclidean hulls of non-separable double cyclic codes over $\mathbb{Z}_2$ for any block lengths $r$ and $s$, completing the work that previous studies left unfinished.\par
The remainder of this paper is organized as follows. Section 2 reviews the preliminary concepts and fundamental algebraic structures of double cyclic codes. In Section 3, we determine the explicit generator polynomials for the Euclidean hulls of these codes and discuss the necessary and sufficient conditions for them to be linear complementary dual (LCD) codes. Section 4 provides concrete computational examples to support and validate our theoretical findings. Finally, Section 5 concludes the paper.

\section{Preliminaries}
In this section, we review the fundamental concepts and basic results concerning cyclic codes and double cyclic codes over $\mathbb{Z}_2$. For further details regarding double cyclic codes over $\mathbb{Z}_2$, we refer the reader to \cite{BCT18}.
\subsection{Cyclic Codes} A cyclic code $\mathcal C$ of length $n$ over $\mathbb{Z}_2$ is a linear code of length $n$ over $\mathbb{Z}_2$ with the property that $(c_0, c_1, \dots, c_{n-1}) \in \mathcal C$ implies $(c_{n-1}, c_0, c_1, \dots, c_{n-2}) \in \mathcal C$. \par
Let ${\bf u}=(u_0,u_1,\ldots, u_{n-1})$ and ${\bf v}=(v_0,v_1,\ldots, v_{n-1})$ be any two element in $\mathbb Z_2^n$ then their Euclidean inner product define as ${\bf u}\cdot{\bf v}=\sum_{i=0}^{n-1}u_iv_i$. The Euclidean dual $\mathcal C^\perp$ of $\mathcal C$ is defined as;
$$\mathcal C^\perp = \{ {\bf u} \in \mathbb{Z}_2^n \mid {\bf u}\cdot {\bf v} = 0 \text{ for all } {\bf v} \in \mathcal C \}.$$
 The Euclidean hull of $\mathcal C$ is defined as 
$$\hull = \mathcal C \cap \mathcal C^\perp.$$
Moreover, if $\hull=\{0\}$, then $\mathcal C$ is called linear complementary dual (LCD) code.\par
For any polynomial $a(x) = \sum_{i=0}^{k} a_i x^i \in \mathbb{Z}_2[x]$ of degree $k$ with $a_0 \neq 0$, its reciprocal polynomial is given by $a^*(x) = x^k a(1/x) = \sum_{i=0}^{k} a_{k-i} x^i$. Clearly, $\deg(a(x)) = \deg(a^*(x))$ and $(a^*)^*(x) = a(x)$. A polynomial $a(x)$ is called self-reciprocal if $a^*(x) = a(x)$. 
\begin{theorem}\label{theo-2.1}\cite{HP03}
Let $\mathcal C_1 = \langle g_1(x) \rangle$ and $\mathcal C_2 = \langle g_2(x) \rangle$ be cyclic codes of length $n$ over $\mathbb{Z}_2$. Then  $\mathcal C_1 \cap \mathcal C_2=\langle\lcm(g_1(x), g_2(x))\rangle$. 
\end{theorem}
\subsection{$\mathbb Z_2$-Double Cyclic Codes}
\begin{definition}\em
Let $r$ and $s$ be positive integers, and let $\mathcal C$ be a binary linear code of  length $n=r+s$. The code $\mathcal C$ is called a double cyclic code over $\mathbb{Z}_2$  of block length $(r,s)$ over if it is invariant under the simultaneous cyclic shift of its coordinates, i.e.,
\[
(u_0, u_1, \dots, u_{r-1} \mid v_0, v_1, \dots, v_{s-1}) \in \mathcal C,
\]
 implies 
\[
(u_{r-1}, u_0, \dots, u_{r-2} \mid v_{s-1}, v_0, \dots, v_{s-2}) \in \mathcal C.
\]
\end{definition}

Let $R$ and $S$ denote the index sets corresponding to the first $r$ coordinates and the last $s$ coordinates of $\mathcal C$, respectively, so that $|R| = r$ and $|S| = s$. By restricting the codewords of $\mathcal C$ to the coordinate sets $R$ and $S$, we obtain the punctured codes, denoted by $\mathcal C_R$ and $\mathcal C_S$. A double cyclic code $C$ is said to be separable if it can be expressed as the direct product of its punctured codes, namely, $C = C_R \times C_S$.\par
 Let $\mathfrak R_{r,s} = \mathbb{Z}_2[x]/\langle x^r - 1 \rangle \times \mathbb{Z}_2[x]/\langle x^s - 1 \rangle$.  Define the canonical bijection $\phi : \mathbb{Z}_2^r \times \mathbb{Z}_2^s \to \mathfrak R_{r,s}$ by
\[
\phi(u_0, u_1, \dots, u_{r-1} \mid v_0, v_1, \dots, v_{s-1}) = (u(x) \mid v(x)),
\]
where $u(x) = \sum_{i=0}^{r-1} u_i x^i$ and $v(x) = \sum_{j=0}^{s-1} v_j x^j$. 
The map $\phi$ is a $\mathbb{Z}_2$-module isomorphism. Under standard polynomial addition and scalar multiplication by elements of $\mathbb{Z}_2[x]$, the space $\mathfrak R_{r,s}$ forms a $\mathbb{Z}_2[x]$-module. Consequently, a linear code $\mathcal C$ is a double cyclic code  of block length $(r,s)$ over $\mathbb{Z}_2$ if and only if its polynomial image $\phi(\mathcal C)$ is a $\mathbb{Z}_2[x]$-submodule of $\mathfrak R_{r,s}$.


\begin{theorem}\label{theo-2.3}\cite{BCT18}
Let $\mathcal C$ be a double cyclic code in  $\mathfrak R_{r,s}$. Then 
$$\mathcal C=\langle (b(x)\mid 0), (l(x)\mid a(x))\rangle,$$
where $a(x), b(x), l(x)\in\mathbb{Z}_2[x]$, with $a(x)$ and $b(x)$ being monic polynomials satisfying $b(x)\mid (x^r-1)$ and $a(x)\mid (x^s-1)$. Moreover, $\deg(l(x))<\deg(b(x))$ and $b(x)\mid \frac{x^s-1}{a(x)}l(x)$ and dimension of $\mathcal C$ is $r+s-\deg(b(x))-\deg(a(x))$ .
\end{theorem}
\begin{corollary}\label{cor-2.4}
    Let $\mathcal C$ be a double cyclic code in  $\mathfrak R_{r,s}$. Then $\mathcal C$ is separable if and only if $l(x) = 0$. 
\end{corollary}
Throughout this paper, let $g_{b,l}(x)=\gcd(b(x),l(x))$. We define 
\begin{equation}
D(x)=\frac{b(x)}{g_{b,l}(x)} \quad \text{and} \quad \widetilde{l}(x)=\frac{l(x)}{g_{b,l}(x)}.
\end{equation}
Since $\gcd(D(x),\widetilde{l}(x))=1$, condition  $b(x)\mid \frac{x^s-1}{a(x)}l(x)$ of  Theorem \ref{theo-2.3} is equivalent to $D(x)\mid \frac{x^s-1}{a(x)}$.

\begin{theorem}\cite{BCT18}\label{theo-2.5}
Let $\mathcal C=\langle (b(x)\mid 0),(lx)\mid a(x))\rangle$ be a non-separable double cyclic code in $\mathfrak R_{r,s}$, and let $\mathcal C^\perp=\langle (\overline{b}(x)\mid 0),(\overline{l}(x)\mid \overline{a}(x))\rangle$ be its dual. Then
\begin{equation}
\overline{b}(x)=\frac{x^r-1}{g_{b,l}^*(x)},
\end{equation}
\begin{equation}
\overline{a}(x)=\frac{(x^s-1)g_{b,l}^*(x)}{a^*(x)b^*(x)}
=\frac{x^s-1}{a^*(x)D^*(x)},
\end{equation}
and
\begin{equation}
\overline{l}(x)=\frac{x^r-1}{b^*(x)}\lambda(x),
\end{equation}
where $m=\operatorname{lcm}(r,s)$
\begin{equation}
\lambda(x)\equiv x^{m-\deg(a(x))+\deg(l(x))}
\big(\widetilde{l}^*(x)\big)^{-1}\pmod{D^*(x)}.
\end{equation}
\end{theorem}
\section{Generators of Hulls of  Double Cyclic
Codes}
Recently, Gao et al. \cite{GWF23} investigated  the generator polynomials  of the Euclidean hulls of separable and non-separable double cyclic codes over $\mathbb{Z}_2$, but restricted their analysis of non-separable codes strictly to coprime block lengths. In this paper, we overcome this limitation by completely characterizing the Euclidean hulls of these codes for arbitrary $r$ and $s$. Consequently, our proposed framework fully generalizes their findings, demonstrating that Gao et al.'s results for both separable and non-separable hulls naturally emerge as special cases of our unified construction.\par

Let $\pi_r : \mathfrak R_{r,s} \to \frac{\mathbb{Z}_2[x]}{\langle x^r - 1 \rangle}$   and $\pi_s : \mathfrak R_{r,s} \to \frac{\mathbb{Z}_2[x]}{\langle x^s - 1 \rangle}$  denote the projection maps  on first $r$ and last $s$ coordinates, respectively. Then $\pi_r(c(x),c'(x))=c(x)$ and   $\pi_s(c(x),c'(x))=c'(x)$, where $(c(x),c'(x))\in \mathfrak R_{r,s}$. From Theorems \ref{theo-2.3} and \ref{theo-2.5}, we have   $$\pi_s(\mathcal C) = \langle a(x) \rangle \quad \text{and} \quad \pi_s(\mathcal C^\perp) = \langle \bar{a}(x) \rangle = \left\langle \frac{x^s - 1}{a^*(x)D^*(x)} \right\rangle.$$
Define \begin{equation}\label{eq-6}
    A_0(x)= \lcm (a(x), \bar{a}(x)).
\end{equation}  Consequently, $\pi_s(\mathcal C) \cap \pi_s(\mathcal C^\perp) = \langle A_0(x) \rangle.$\par
As established in [4], when $\gcd(r,s) = 1$, the projection of the hull $\pi_s(\hull)$ precisely coincides with the intersection of the individual projections, yielding $\pi_s(\mathcal C) \cap \pi_s(\mathcal C^\perp) = \langle A_0(x) \rangle$.
\begin{proposition}\cite{GWF23}\label{prop-3.1}
 Let $\mathcal{C}$, $\mathcal{C}^\perp$,  $\pi_r$ and $\pi_s$ be as define previously. Then 
   $$\pi_r(C \cap C^\perp) \subseteq \pi_r(C) \cap \pi_r(C^\perp)\mbox{~and~}\pi_s(C \cap C^\perp) \subseteq \pi_s(C) \cap \pi_s(C^\perp).$$
   
\end{proposition}
As a direct consequence of the above proposition,  $\pi_s\hull$ is constrained to be a subcode of $\langle A_0(x) \rangle$. This relationship may form a proper subcode since the cross-coordinate coupling dictates that not every polynomial generated by $A_0(x)$ will have a corresponding first coordinate residing in the intersection $C \cap C^\perp$.\par
The following lemma characterizes the exact lifting condition required to find such a compatible first coordinate.
\begin{lemma}\label{lemma-3.2}
Let $\mathcal C = \langle (b(x) \mid 0), (l(x) \mid a(x)) \rangle$ be a double cyclic code in $\mathfrak R_{r,s}$, and  $\mathcal C^\perp = \langle (\bar{b}(x) \mid 0), (\bar{l}(x) \mid \bar{a}(x)) \rangle$ be its dual. Define
\begin{equation}\label{eq-7}
    G(x) = \gcd(b(x), \bar{b}(x)) = \gcd\left(b(x), \frac{x^r - 1}{g^*_{b,l}(x)}\right),
\end{equation}
\begin{equation}\label{eq-8}
    E(x) = \frac{A_0(x)}{a(x)}l(x) + \frac{A_0(x)}{\bar{a}(x)}\bar{l}(x) \in \mathbb{Z}_2[x].
\end{equation}
For any polynomial $v(x) \in \mathbb{Z}_2[x]$, there exists $u(x) \in \mathbb{Z}_2[x]$ such that $(u(x) \mid v(x)) \in \hull$ if and only if $v(x) = k(x)A_0(x)$ for some $k(x) \in \mathbb{Z}_2[x]$ such that $$k(x)E(x) \equiv 0 \pmod{G(x)}.$$
\end{lemma}
\begin{proof}
Suppose that $(u(x) \mid v(x)) \in \hull$. Then, $v(x)\in\pi_s(\mathcal C) \cap \pi_s(\mathcal C^\perp) = \langle A_0(x) \rangle$. Consequently, $v(x)$ can be written as $v(x) = k(x)A_0(x)$ for some polynomial $k(x) \in \mathbb{Z}_2[x]$. Since $(u(x) \mid v(x)) \in \mathcal C$ and $\frac{k(x)A_0(x)}{a(x)}(l(x)\mid a(x))\in \mathcal C$, we have  
\[(u(x) \mid v(x)) -\frac{k(x)A_0(x)}{a(x)}(l(x)\mid a(x))=(u(x) +\frac{k(x)A_0(x)}{a(x)}l(x)\mid 0)\in \mathcal C.\]
This relation holds if and only if the first coordinate is a multiple of $b(x)$. Thus 
\begin{equation}\label{eq-9}
u(x) = f_1(x)b(x) + \dfrac{k(x)A_0(x)}{a(x)}l(x)
\end{equation}for some $f_1(x) \in \mathbb{Z}_2[x]$.
Applying the same argument to the dual code $\mathcal C^\perp$, we deduce that $(u(x) \mid v(x)) \in \mathcal C^\perp$ if and only if
\begin{equation}\label{eq-10}
u(x) = f_2(x)\bar{b}(x) +\dfrac {k(x)A_0(x)}{\bar{a}(x)}\bar{l}(x),
\end{equation}for some $f_2(x) \in \mathbb{Z}_2[x]$.
Equating Eq.\eqref{eq-9} and Eq.\eqref{eq-10},  we get a common polynomial $u(x)$ exists if and only if we can find $f_1(x), f_2(x) \in \mathbb{Z}_2[x]$ that satisfy
$$f_1(x)b(x) + f_2(x)\bar{b}(x) = k(x)\left( \frac{A_0(x)}{a(x)}l(x) + \frac{A_0(x)}{\bar{a}(x)}\bar{l}(x) \right) = k(x)E(x).$$
Since  $\langle b(x), \bar{b}(x) \rangle=\langle G(x)\rangle.$  Therefore, such polynomials $f_1(x)$ and $f_2(x)$ exist if and only if $G(x)$ divides $k(x)E(x)$, which is precisely equivalent to 
$$k(x)E(x) \equiv 0 \pmod{G(x)}.$$ This completes the proof.
\end{proof}
Now, we give the generator polynomials of Hulls of  double cyclic code in $\mathfrak R_{r,s}$.  Let $A_0(x)$, $G(x)$, and $E(x)$ be defined as in Eqs. \eqref{eq-6}-\eqref{eq-8}, and define 
\begin{equation}\label{eq-11}
    \tilde{ k}(x) = \frac{G(x)}{\gcd(G(x), E(x))}.
\end{equation}Let $\alpha(x), \beta(x) \in \mathbb{Z}_2[x]$ be Bézout coefficients satisfying the identity $\alpha(x)b(x) + \beta(x)\bar{b}(x) = G(x)$. We know that Hulls of double cyclic code in $\mathfrak R_{r,s}$ is  also a double cyclic code in $\mathfrak R_{r,s}$. 
\begin{theorem}\label{theo-3.3}
 Let $\mathcal C = \langle (b(x) \mid 0), (l(x) \mid a(x)) \rangle$ be a double cyclic code in $\mathfrak R_{r,s}$, and  $\mathcal C^\perp = \langle (\bar{b}(x) \mid 0), (\bar{l}(x) \mid \bar{a}(x)) \rangle$ be its Euclidean dual. Then $$\hull = \langle (B(x) \mid 0), (L(x) \mid A(x)) \rangle,$$where 
 \begin{equation}\label{eq-12}
     B(x) = \lcm(b(x), \bar{b}(x)) = \lcm\left(b(x), \frac{x^r - 1}{g^*_{b,l}(x)}\right),
 \end{equation}
 \begin{equation}\label{eq-13}
     A(x) = \tilde{ k}(x)A_0(x),
 \end{equation}
 \begin{equation}\label{eq-14}
    L(x) = \left( \tilde{ k}(x)\frac{A_0(x)}{a(x)}l(x) + \alpha(x)\frac{\tilde{ k}(x)E(x)}{G(x)}b(x) \right) \bmod B(x).
 \end{equation} Furthermore, $$\dim(\hull) = r + s - \deg(B(x)) - \deg(A(x)).$$
\end{theorem}

\begin{proof}
First, we establish the generator $B(x)$ for the subcode of $\hull$ consisting of codewords with a zero second coordinate, given by $\{ (u(x) \mid 0) \in \mathcal C \cap \mathcal C^\perp \}$. By Proposition \ref{prop-3.1} the generator $\pi_r(\hull)\subseteq \langle \lcm(b(x), \bar b(x))\rangle$. Let $(u(x) \mid 0) \in \mathcal C \cap \mathcal C^\perp $. Then $(u(x) \mid 0) \in \mathcal C$ it holds if and only if $b(x) \mid u(x)$, and $(u(x) \mid 0) \in \mathcal C^\perp$ if and only if $\bar{b}(x) \mid u(x)$. So, we  get  that $ \lcm(b(x), \bar b(x))|u(x).$ This implies that $\langle \lcm(b(x), \bar b(x))\rangle\subseteq \pi_r(\hull).$ Therefore,
$$B(x) = \text{lcm}(b(x), \bar{b}(x)).$$
Now, we determine the second generator component $A(x)$. For any $(u(x) \mid v(x)) \in \hull$, the second coordinate $v(x)$ must belong to both $\pi_s(\mathcal C) = \langle a(x) \rangle$ and $\pi_s(\mathcal C^\perp) = \langle \bar{a}(x) \rangle$. Thus, $v(x)$ must be a multiple of $A_0(x) = \text{lcm}(a(x), \bar{a}(x))$, implying
$$v(x) = k(x)A_0(x),$$
where $k(x) \in \mathbb{Z}_2[x]$ satisfies the divisibility condition $G(x) \mid k(x)E(x)$ by Lemma \ref{lemma-3.2}. This condition is algebraically equivalent to stating that $k(x)$ is a multiple of $\tilde{ k}(x) = \frac{G(x)}{\gcd(G(x), E(x))}$. Consequently, we can write $k(x) = m(x)\tilde{ k}(x)$ for some polynomial $m(x) \in \mathbb{Z}_2[x]$, which yields:
$$v(x) = m(x)\tilde{ k}(x)A_0(x) = m(x)A(x).$$
This demonstrates that every element in $\pi_s(\hull)$ is a multiple of $A(x)$, establishing the inclusion $\pi_s(\hull) \subseteq \langle A(x) \rangle$.

To prove the reverse containment $\langle A(x) \rangle \subseteq \pi_s(\hull)$, we must show that for any multiple $m(x)A(x)$ (and in particular for $m(x) = 1$, corresponding to $A(x)$ itself), there exists a compatible first coordinate $L_0(x)$ such that $(L_0(x) \mid A(x)) \in \mathcal C \cap \mathcal C^\perp$. By the definition of $\tilde{ k}(x)$, we have $\tilde{ k}(x)E(x) = q(x)G(x)$, where $q(x) = \frac{\tilde{ k}(x)E(x)}{G(x)} = \frac{E(x)}{\gcd(G(x), E(x))}$.
Using Bézout's identity, let  $\alpha(x)b(x) + \beta(x)\bar{b}(x) = G(x)$. Multiplying both sides by $q(x)$ yields:
$$q(x)\big(\alpha(x)b(x) + \beta(x)\bar{b}(x)\big) = q(x)G(x) \implies q(x)\alpha(x)b(x) + q(x)\beta(x)\bar{b}(x) = \tilde{ k}(x)E(x).$$
Substituting, $E(x) = \frac{A_0(x)}{a(x)}l(x) + \frac{A_0(x)}{\bar{a}(x)}\bar{l}(x)$, we obtain that 
$$q(x)\alpha(x)b(x) + q(x)\beta(x)\bar{b}(x) = \frac{\tilde{ k}(x)A_0(x)}{a(x)}l(x) + \frac{\tilde{ k}(x)A_0(x)}{\bar{a}(x)}\bar{l}(x).$$
By rearranges, we have 
$$\frac{\tilde{ k}(x)A_0(x)}{a(x)}l(x) + q(x)\alpha(x)b(x) = \frac{\tilde{ k}(x)A_0(x)}{\bar{a}(x)}\bar{l}(x) + q(x)\beta(x)\bar{b}(x).$$
Now, define
$$L_0(x) = \frac{\tilde{ k}(x)A_0(x)}{a(x)}l(x) + q(x)\alpha(x)b(x) = \frac{\tilde{ k}(x)A_0(x)}{\bar{a}(x)}\bar{l}(x) + q(x)\beta(x)\bar{b}(x).$$
Then, for any $m(x) \in \mathbb{Z}_2[x]$, setting $u(x) = m(x)L_0(x) \pmod{x^r - 1}$ provides a compatible first coordinate for $v(x) = m(x)A(x)$ such that
$$\begin{aligned}
(u(x) \mid v(x)) &= m(x)q(x)\alpha(x)(b(x) \mid 0) + m(x)\frac{\tilde{ k}(x)A_0(x)}{a(x)}(l(x) \mid a(x)) \in \mathcal C, \\
(u(x) \mid v(x)) &= m(x)q(x)\beta(x)(\bar{b}(x) \mid 0) + m(x)\frac{\tilde{ k}(x)A_0(x)}{\bar{a}(x)}(\bar{l}(x) \mid \bar{a}(x)) \in \mathcal C^\perp.
\end{aligned}$$
Thus, $(L_0(x) \mid A(x)) \in \mathcal C \cap \mathcal C^\perp = \hull$.
 Let $\deg (L(x))>\deg(B(x))$, applying the Euclidean division algorithm, we get  $L_0(x) = t(x)B(x) + L(x)$, where $\deg(B(x))>\deg(L(x))$.  Now, 
$$(L(x) \mid A(x)) = (L_0(x) \mid A(x)) + t(x)(B(x) \mid 0) \in \hull,$$
which establishes the equality of submodules $\langle (B(x) \mid 0), (L_0(x) \mid A(x)) \rangle = \langle (B(x) \mid 0), (L(x) \mid A(x)) \rangle$. This completes the proof.
\end{proof}
In the following corollary, we establish the conditions under which the Euclidean hull of a non-separable double cyclic code degenerates into a separable double cyclic code.
\begin{theorem}\label{theo-3.4}
Let $\mathcal C=\langle(b(x)\mid 0),(l(x)\mid a(x))\rangle$
be a non-separable double cyclic code in $\mathcal R_{r,s}$.
Then $\hull$ is a separable double cyclic
code if and only if
\[
A(x)=\lcm\left(
a(x)D(x),\frac{x^s-1}{a^*(x)}
\right),
\]
where $A(x)=\widetilde{k}(x)A_0(x)$.
\end{theorem}

\begin{proof}
By Theorem~\ref{theo-3.3}, we have
\[
\hull=
\langle(B(x)\mid 0),(L(x)\mid A(x))\rangle.
\]
By Corollary~2.4, $\hull$ is separable
if and only if $L(x)=0$, or equivalently,
$(0\mid A(x))\in \mathcal C\cap \mathcal C^\perp$.
Observe that $(0\mid A(x))\in \mathcal C$ if and only if
\[
b(x)\mid \frac{A(x)}{a(x)}l(x).
\]
Since $b(x)=g_{b,l}(x)D(x)$,
$l(x)=g_{b,l}(x)\widetilde{l}(x)$, and
$\gcd(D(x),\widetilde{l}(x))=1$, this is equivalent to
\[
a(x)D(x)\mid A(x).
\]
Similarly, $(0\mid A(x))\in \mathcal C^\perp$ if and only if
\[
\overline{b}(x)\mid
\frac{A(x)}{\overline{a}(x)}\overline{l}(x).
\]
By Theorem~2.5, we have
\[
\frac{\overline{b}(x)}
{\gcd(\overline{b}(x),\overline{l}(x))}
=D^*(x).
\]
Consequently, the above divisibility condition is
equivalent to
\[
\overline{a}(x)D^*(x)
=\frac{x^s-1}{a^*(x)}\mid A(x).
\]
Therefore, $\hull$ is separable
if and only if
\[
\lcm\left(
a(x)D(x),\frac{x^s-1}{a^*(x)}
\right)\mid A(x).
\]
Now, put
\[
M(x)=\lcm\left(
a(x)D(x),\frac{x^s-1}{a^*(x)}
\right).
\]
By the preceding arguments, $(0\mid M(x))$ belongs
to both $C$ and $C^\perp$, and hence
$(0\mid M(x))\in\lcm$.
Since $\pi_s(\hull=\langle A(x)\rangle$,
we obtain $A(x)\mid M(x)$.
Thus, $\hull$ is separable if and only if
\[
A(x)=\lcm\left(
a(x)D(x),\frac{x^s-1}{a^*(x)}
\right).
\]
This completes the proof.
\end{proof}

Following corollary is a direct consequence of Theorem \ref{theo-3.3}, if we take $l(x)=0.$ 
\begin{corollary}\label{corro-3.5}
 Let $\mathcal C = \langle (b(x) \mid 0), (0 \mid a(x)) \rangle$ be a  separable double cyclic code in $\mathfrak R_{r,s}$. Then $$\hull = \langle (B'(x) \mid 0), (0 \mid A'(x)) \rangle,$$where 
 \begin{equation*}
     B'(x) = \lcm(b(x), \bar{b}(x)) = \lcm\left(b(x), \frac{x^r - 1}{b^*(x)}\right),
 \end{equation*}
 \begin{equation*}
     A'(x) = \lcm\left(a(x), \frac{x^s-1}{a^*(x)}\right).
 \end{equation*}

\end{corollary}
We now establish the necessary and sufficient algebraic conditions for a non-separable double cyclic code to be a linear complementary dual (LCD) code.

\begin{theorem}\label{theo-3.6}
Let $\mathcal C = \langle (b(x) \mid 0), (l(x) \mid a(x)) \rangle$ be a non-separable double cyclic code in $\mathfrak R_{r,s}$. Then $\mathcal C$ (equivalently $\mathcal C^\perp$) is a linear complementary dual (LCD) code  if and only if the following three conditions hold simultaneously
\begin{enumerate}
    \item[\textup{(i)}] $\gcd\!\left(g_{b,l}^*(x),\, \frac{x^r - 1}{b(x)}\right) = 1$, or equivalently, $B(x) = \mathrm{lcm}(b(x), \bar{b}(x)) = x^r - 1$;
    and
    \item[\textup{(ii)}] $ \frac{G(x)}{\gcd(G(x), E(x))} = \frac{x^s - 1}{A_0(x)}$.
\end{enumerate}
Furthermore, under these conditions $L(x) = 0$.
\end{theorem}

\begin{proof}
By definition, $\mathcal C$ is an LCD code if and only if $\hull = \{\mathbf{0}\}$, which is equivalent to the condition $\dim(\hull) = 0$. According to Theorem~\ref{theo-3.3},  $\hull = \langle (B(x) \mid 0), (L(x) \mid A(x)) \rangle$, and 
\[
\dim(\hull) = r+s- \deg(B(x)) - \deg(A(x)).
\]
Since $B(x)$ and $A(x)$ are monic divisors of $x^r - 1$ and $x^s - 1$, respectively, we have $\dim(\hull) = 0$ if and only if $B(x) = x^r - 1$ and $A(x) = x^s - 1$. Notice that when $B(x) = x^r - 1$ and $A(x) = x^s - 1$, the condition $B(x) \mid \frac{x^s - 1}{A(x)}L(x)$ reduces to $B(x) \mid L(x)$. Because $\deg(L(x)) < \deg(B(x)) = r$, this divisibility forces $L(x) = 0$.
First, we establish Condition~(i) by characterizing the requirement $B(x) = x^r - 1$. Since $B(x) = \mathrm{lcm}(b(x), \bar{b}(x))$, we can write
\[
B(x) = \mathrm{lcm}\!\left(b(x),\, \frac{x^r - 1}{g_{b,l}^*(x)}\right) = \frac{b(x)\frac{x^r - 1}{g_{b,l}^*(x)}}{\gcd\!\left(b(x),\, \frac{x^r - 1}{g_{b,l}^*(x)}\right)} = \frac{x^r - 1}{\gcd\!\left(\frac{x^r - 1}{b(x)},\, g_{b,l}^*(x)\right)}.
\]
Thus, $B(x) = x^r - 1$ holds if and only if $\gcd\!\left(g_{b,l}^*(x),\, \frac{x^r - 1}{b(x)}\right) = 1$.
Next, we establish Conditions~(ii) and~(iii) by analyzing the requirement $A(x) = x^s - 1$. By Theorem~\ref{theo-3.3}, we have $A(x) = \tilde{ k}(x)A_0(x)$, where $\tilde{ k}(x) = \frac{G(x)}{\gcd(G(x), E(x))}$. Consequently, $A(x) = x^s - 1$ if and only if
\[
\tilde{ k}(x) = \frac{G(x)}{\gcd(G(x), E(x))} = \frac{x^s - 1}{A_0(x)}.
\]
 This directly yields Conditions~(ii) and completes the proof.
\end{proof}
\begin{corollary}\label{corr-3.7}
Let $\mathcal{C} = \langle(b(x) \mid 0), (l(x) \mid a(x))\rangle$ be a non-separable double cyclic code in $\mathfrak{R}_{r,s}$ such that $r$ and $s$ are odd and $\gcd(r,s) = 1$. Then $\mathcal{C}$ can never be a linear complementary dual (LCD) code.
\end{corollary}

\begin{proof}
Since $r$ and $s$ are coprime odd positive integers, it follows directly from Lemma~4.2 and Corollary~4.3 of  \cite{GWF23} that $\mathcal C$ and its Euclidean dual $\mathcal C^\perp$ take the explicit canonical forms
\[
\mathcal C = \big\langle ((x - 1)Q_1(x) \mid 0),\, (Q_1(x) \mid a(x)) \big\rangle
\]
and
\[
\mathcal C^\perp = \left\langle \left(\frac{x^r - 1}{Q_1^*(x)} \;\middle|\; 0\right),\, \left(\frac{x^r - 1}{(x - 1)Q_1^*(x)} \;\middle|\; \frac{x^s - 1}{(x - 1)a^*(x)}\right) \right\rangle,
\]
where $Q_1(x) \mid (x^r - 1)$, $a(x) \mid (x^s - 1)$, $(x - 1) \nmid Q_1(x)$, and $(x - 1) \nmid a(x)$. 

In the notation of Theorem~3.2, we have $b(x) = (x - 1)Q_1(x)$, $l(x) = Q_1(x)$, $\bar{b}(x) = \frac{x^r - 1}{Q_1^*(x)}$, $\bar{l}(x) = \frac{x^r - 1}{(x - 1)Q_1^*(x)}$, and $\bar{a}(x) = \frac{x^s - 1}{(x - 1)a^*(x)}$. Since $r$ and $s$ are odd, $x - 1$ divides both $x^r - 1$ and $x^s - 1$ with multiplicity~$1$. Consequently:
\begin{enumerate}
    \item[(1)] Because $(x - 1) \nmid Q_1(x)$ and $(x - 1) \nmid Q_1^*(x)$, the factor $x - 1$ divides $G(x) = \gcd(b(x), \bar{b}(x))$ with multiplicity~$1$.
    \item[(2)] Because $(x - 1) \nmid a(x)$ and $(x - 1) \nmid \bar{a}(x)$, we have $(x - 1) \nmid A_0(x) = \mathrm{lcm}(a(x), \bar{a}(x))$.
    \item[(3)] Because $(x - 1) \nmid l(x)$ and $(x - 1) \nmid \bar{l}(x)$, evaluating at $x = 1$ over $\mathbb{Z}_2$ gives $l(1) = \bar{l}(1) = \frac{A_0(1)}{a(1)} = \frac{A_0(1)}{\bar{a}(1)} = 1$.
\end{enumerate}

Evaluating the compatibility polynomial $E(x) = \frac{A_0(x)}{a(x)}l(x) + \frac{A_0(x)}{\bar{a}(x)}\bar{l}(x)$ at $x = 1$ over $\mathbb{Z}_2$ yields
\[
E(1) = 1 \cdot 1 + 1 \cdot 1 = 0 \pmod 2,
\]
so $(x - 1) \mid E(x)$. Since $x - 1$ divides $G(x)$ with multiplicity~$1$, we have $(x - 1) \mid \gcd(G(x), E(x))$, which implies $(x - 1) \nmid \tilde{ k}(x) = \frac{G(x)}{\gcd(G(x), E(x))}$. Together with $(x - 1) \nmid A_0(x)$, this gives $(x - 1) \nmid A(x) = \tilde{ k}(x)A_0(x)$. Hence $A(x) \neq x^s - 1$ which implies $\deg(A(x)) \le s - 1$ and $\dim(\hull) \ge s - \deg(A(x)) \ge 1$. Therefore, $\mathcal C$ can never be an LCD code.
\end{proof}
\begin{note}\em
In \cite[Theorem 4.9]{GWF23}, Gao et al. provided a dimension formula for the hull of a non-separable double cyclic code when $r$ and $s$ are odd coprimes. Their formula consists of a constant term of $1$ added to a non-negative function dependent on $r$ and $s$, which guarantees that the hull dimension is always at least $1$. Since the dimension is strictly greater than zero, the code can never be an LCD code. This observation is entirely consistent with, and validates, our Corollary \ref{corr-3.7}.
\end{note}

\section{Worked Example}
To validate our theoretical results, we now provide concrete examples of non-separable double cyclic codes and explicitly compute their Euclidean hulls. All examples presented in this paper were verified using the Magma computer algebra system \cite{BCP97}.
\begin{example}\em\label{ex-4.1}
Let $r = 3$ and $s = 9$, so $\gcd(r,s) = 3 > 1$. Let $x^3 - 1 = (x + 1)(x^2 + x + 1)$ and $x^9 - 1 = (x + 1)(x^2 + x + 1)(x^6 + x^3 + 1)$. Consider  a nonseparable double cyclic code $\mathcal C = \langle (b(x) \mid 0), (l(x) \mid a(x)), \rangle$  of length $(3,9)$, where
\[
b(x) = x^2 + x + 1, \quad l(x) = 1, \quad a(x) = x^6 + x^3 + 1.
\]
Suppose $\mathcal C^\perp = \langle (\bar{b}(x) \mid 0), (\bar{l}(x) \mid \bar{a}(x)) \rangle$, observe that $g_{b,l}(x) = \gcd(x^2 + x + 1, 1) = 1$, $b^*(x) = b(x)$, and $a^*(x) = a(x)$. By Theorem 2.5, we obtain
\[
\bar{b}(x) = \frac{x^3 - 1}{1} = x^3 - 1, \quad \bar{a}(x) = \frac{x^9 - 1}{(x^6 + x^3 + 1)(x^2 + x + 1)} = x + 1.
\]
Moreover, $\lambda(x) = x^{9 - 6 + 0} \cdot 1^{-1} \bmod (x^2 + x + 1) = 1$, which gives
\[
\bar{l}(x) = \frac{x^3 - 1}{x^2 + x + 1}\lambda(x) = x + 1.
\]
Now 
\begin{align*}
B(x) &= \mathrm{lcm}(b(x), \bar{b}(x)) = \mathrm{lcm}(x^2 + x + 1, x^3 - 1) = x^3 - 1, \\
A_0(x) &= \mathrm{lcm}(a(x), \bar{a}(x)) = \mathrm{lcm}(x^6 + x^3 + 1, x + 1) = (x + 1)(x^6 + x^3 + 1), \\
G(x) &= \gcd(b(x), \bar{b}(x)) = \gcd(x^2 + x + 1, x^3 - 1) = x^2 + x + 1.
\end{align*}
Since $\frac{A_0(x)}{a(x)} = x + 1$ and $\frac{A_0(x)}{\bar{a}(x)} = x^6 + x^3 + 1$, the compatibility polynomial is
\[
E(x) = (x + 1)\cdot 1 + (x^6 + x^3 + 1)(x + 1) = x^3(x + 1)(x^3 - 1) = x^3(x + 1)^2(x^2 + x + 1).
\]
Since $G(x) = x^2 + x + 1$ divides $E(x)$, we have $\gcd(G(x), E(x)) = G(x)$ and $\tilde{ k}(x) = 1$. Hence,
\[
A(x) = \tilde{ k}(x)A_0(x) = (x + 1)(x^6 + x^3 + 1) = x^7 + x^6 + x^4 + x^3 + x + 1.
\]
Taking Bézout coefficients $\alpha(x) = 1$ and $\beta(x) = 0$ for $G(x) = b(x)$, we have $q(x) = \frac{\tilde{ k}(x)E(x)}{G(x)} = x^3(x + 1)^2$, which yields
\[
L_0(x) = (x + 1) + x^3(x + 1)^2(x^2 + x + 1) = (x + 1) + x^3(x + 1)(x^3 - 1).
\]
Reducing modulo $B(x) = x^3 - 1$ gives $L(x) = L_0(x) \bmod (x^3 - 1) = x + 1 \neq 0$. Thus, $\hull = \langle (x^3 - 1 \mid 0), (x + 1 \mid x^7 + x^6 + x^4 + x^3 + x + 1) \rangle$ is a non-separable double cyclic code of dimension $\dim(\hull) = 2$.
\end{example}
\begin{example}\cite[Example 1]{GWF23}\em
Let $r = 3$ and $s = 17$,so  $\gcd(3, 17) = 1$ with total length $n = r + s = 20$.  Consider the non-separable double cyclic code $\mathcal C = \langle (b(x) \mid 0), (l(x) \mid a(x)) \rangle \subseteq R_{3,17}$, where  $b(x) = x + 1$, $l(x) = 1$, and $a(x) = \sum_{i=0}^{16} x^i$.  Following Theorem 2.5(2), the Euclidean dual is given by $\mathcal C^\perp = \langle (x^3 - 1 \mid 0), (x^2 + x + 1 \mid 1) \rangle$. Applying Theorem 3.2, we compute $B(x) = \mathrm{lcm}(b(x), \bar{b}(x)) = x^3 - 1$, $A_0(x) = \mathrm{lcm}(a(x), \bar{a}(x)) = a(x)$, and $G(x) = \gcd(b(x), \bar{b}(x)) = x + 1$. The compatibility polynomial satisfies $E(x) = 1 + a(x)(x^2 + x + 1)$, which is a multiple of $G(x)$. Consequently, $\tilde{ k}(x) = 1$, yielding $A(x) = a(x) = \sum_{i=0}^{16} x^i$. Using Bézout coefficients $\alpha(x) = 1$ and $\beta(x) = 0$ for $G(x) = b(x)$,  $L_0(x) = 1 + E(x) = a(x)(x^2 + x + 1)$. Reducing $L_0(x)$ modulo $B(x) = x^3 - 1$ gives
    \[
    L(x) = \big(x^2 \cdot (x^2 + x + 1)\big) \bmod (x^3 - 1) = (x^4 + x^3 + x^2) \bmod (x^3 - 1) = x^2 + x + 1 \neq 0.
    \] Thus,  $\hull = \langle (x^3 - 1 \mid 0), (x^2 + x + 1 \mid \sum_{i=0}^{16} x^i) \rangle$. 
\end{example}
\begin{example}\em
Let $r = 5$ and $s = 5$, so $\gcd(r, s) = 5 > 1$. Consider the non-separable double cyclic code $\mathcal{C} = \langle(b(x) \vert{} 0), (l(x) \vert{} a(x))\rangle \subseteq \mathfrak{R}_{5,5}$ with $b(x) = x^4 + x^3 + x^2 + x + 1$, $l(x) = x + 1$, and $a(x) = 1$.Since $g_{b,l}(x) = \gcd(x^4 + x^3 + x^2 + x + 1, x + 1) = 1$ and $D(x) = x^4 + x^3 + x^2 + x + 1$, Theorem 2.5(2) gives $\overline{b}(x) = x^5 - 1$, $\overline{a}(x) = x + 1$, and $\overline{l}(x) = x^4 + x^3 + x^2 + 1$.Applying Theorem 3.3, we have $B(x) = \text{lcm}(b(x), \overline{b}(x)) = x^5 - 1$, $A_0(x) = \text{lcm}(a(x), \overline{a}(x)) = x + 1$, and $G(x) = \gcd(b(x), \overline{b}(x)) = x^4 + x^3 + x^2 + x + 1$. The compatibility polynomial satisfies$$E(x) = \frac{A_0(x)}{a(x)}l(x) + \frac{A_0(x)}{\overline{a}(x)}\overline{l}(x) = (x+1)(x+1) + 1 \cdot (x^4 + x^3 + x^2 + 1) = x^4 + x^3.$$Since $G(x)$ shares no roots with $x^4 + x^3$ over $\mathbb{Z}_2$, we get $\gcd(G(x), E(x)) = 1$, which implies $\tilde{k}(x) = x^4 + x^3 + x^2 + x + 1$.Consequently, $A(x) = \tilde{k}(x)A_0(x) = (x^4 + x^3 + x^2 + x + 1)(x + 1) = x^5 - 1$.
Choosing B'ezout coefficients $\alpha(x) = 1$ and $\beta(x) = 0$ for $G(x) = b(x)$, we obtain $q(x) = \frac{\tilde{k}(x)E(x)}{G(x)} = x^4 + x^3$, which yields$$L_0(x) = \frac{\tilde{k}(x) A_0(x)}{a(x)} l(x) + q(x)\alpha(x)b(x) = (x^5-1)(x+1) + (x^4+x^3)(x^4 + x^3 + x^2 + x + 1)$$$$L_0(x) = x^8 + x^6 + x^5 + x^3 + x + 1.$$Since $B(x) = x^5 - 1$ completely divides $L_0(x)$ (specifically, $L_0(x) = (x^5-1)(x^3+x+1)$), we get $L(x) = L_0(x) \pmod{x^5 - 1} = 0$.Thus, $\text{Hull}(\mathcal{C}) = \langle(x^5 - 1 \vert{} 0), (0 \vert{} x^5 - 1)\rangle = \{0\}$, demonstrating that $\mathcal{C}$ is a non-separable LCD double cyclic code.
\end{example}

\begin{example}\em
Let $r = 7$ and $s = 7$, so that $n = 14$, $\gcd(r,s) = 7 > 1$, and $m = \mathrm{lcm}(7,7) = 7$. Over $\mathbb{Z}_2$, $x^7 - 1 = (x + 1)(x^3 + x + 1)(x^3 + x^2 + 1)$. Consider$\mathcal C = \langle (b(x) \mid 0), (l(x) \mid a(x)) \rangle \subseteq R_{7,7}$, where
\[
b(x) = x^3 + x + 1, \quad l(x) = 1, \quad a(x) = (x + 1)(x^3 + x^2 + 1) = x^4 + x^2 + x + 1.
\]
By Theorem \ref{theo-3.3}, we compute $B(x) = x^7 - 1$, $A_0(x) = a(x)$, and $G(x) = x^3 + x + 1 = b(x)$. The compatibility polynomial is given by
\[
E(x) = 1 \cdot 1 + a(x)\bar{l}(x) = 1 + (x + 1)^2(x^2 + 1)(x^3 + x^2 + 1)(x^3 + x + 1).
\]
Since $G(x) = x^3 + x + 1$ divides $a(x)\bar{l}(x)$, we have $E(x) \equiv 1 \pmod{G(x)}$, which implies $\gcd(G(x), E(x)) = 1$ and $\tilde{ k}(x) = x^3 + x + 1$. Consequently,
\[
A(x) = \tilde{ k}(x)A_0(x) = (x^3 + x + 1)(x + 1)(x^3 + x^2 + 1) = x^7 - 1.
\]
Choosing Bézout coefficients $\alpha(x) = 1$ and $\beta(x) = 0$ for $G(x) = b(x)$, we obtain $q(x) = \frac{\tilde{ k}(x)E(x)}{G(x)} = 1 + a(x)\bar{l}(x)$ and
\[
L_0(x) = (x^3 + x + 1) + \big(1 + a(x)\bar{l}(x)\big)(x^3 + x + 1) = (x^3 + x + 1)a(x)\bar{l}(x) = (x^7 - 1)\bar{l}(x).
\]
Since $B(x) = x^7 - 1$ divides $L_0(x)$, Corollary 3.3 implies $L(x) = L_0(x) \bmod B(x) = 0$. Therefore, $\hull = \langle (x^7 - 1 \mid 0), (0 \mid x^7 - 1) \rangle = \{\mathbf{0}\}$ is a separable code of dimension $0$, and $\mathcal C$ is a non-separable  LCD double cyclic code.
\end{example}

\begin{example}\em\label{ex-4.5}
Let $r = 4$ and $s = 4$, $\gcd(r,s) = 4 > 1$. Consider the non-separable double cyclic code $\mathcal C = \langle (b(x) \mid 0), (l(x) \mid a(x)) \rangle \subseteq R_{4,4}$ with $b(x) = (x + 1)^2$, $l(x) = x + 1$, and $a(x) = (x + 1)^2$. By Theorem\ref{theo-2.5}, the dual code $\mathcal C^\perp$ has generators $\bar{b}(x) = (x + 1)^3$, $\bar{l}(x) = (x + 1)^2$, and $\bar{a}(x) = x + 1$. From Theorem~\ref{theo-3.3}, we compute $B(x) = (x + 1)^3$, $A_0(x) = (x + 1)^2$, $G(x) = (x + 1)^2$, and
\[
E(x) = (x + 1) + (x + 1)(x + 1)^2 = x^2(x + 1).
\]
Consequently , $\gcd(G(x), E(x)) = x + 1$ and $\tilde{ k}(x) = x + 1$, yielding $A(x) = \tilde{ k}(x)A_0(x) = (x + 1)^3$. Choosing Bézout coefficients $\alpha(x) = 1$ and $\beta(x) = 0$ for $G(x) = b(x)$, we obtain $q(x) = \frac{\tilde{ k}(x)E(x)}{G(x)} = x^2$ and
$L_0(x) = (x + 1)^2 + x^2(x + 1)^2 = (x + 1)^4.$
Since $B(x) = (x + 1)^3$ divides $L_0(x) = (x + 1)^4$, Corollary 3.3 implies $L(x) = L_0(x) \bmod B(x) = 0$. Therefore, $\hull = \langle ((x + 1)^3 \mid 0), (0 \mid (x + 1)^3) \rangle$ is a non-trivial separable double cyclic code with  $\dim(\hull) = 2$.
\end{example}
\section{Conclusion}
In this paper, we investigated the Euclidean hulls of double cyclic codes for arbitrary block lengths, successfully resolving the open problem posed by Gao et al.~\cite{GWF23}, whose prior work on the hulls of non-separable double cyclic codes was strictly restricted to coprime block lengths. We established explicit generator polynomials for the hulls of these codes for arbitrary lengths in Theorem~\ref{theo-3.3}. As a result, we showed that the earlier results by Gao et al.\ \cite[Theorems 3.1 and 4.7]{GWF23} are simply special cases of our broader findings in Corollaries~\ref{corro-3.5} and~\ref{corr-3.7}. In Theorem~\ref{theo-3.4}, we provided the exact conditions for when the hull of a non-separable double cyclic code reduces to a separable one.  We also derived the necessary and sufficient conditions for a non-separable double cyclic code to be a LCD code in Theorem~\ref{theo-3.6}. Finally, to validate our theoretical formulations, we provided several computational examples (Examples~\ref{ex-4.1}--\ref{ex-4.5}), all of which were verified using the Magma computer algebra system. \par
Because the polynomial $E(x)$ has non-linear coefficients, enumerating double cyclic codes of arbitrary block lengths with a fixed hull dimension is a  an interesting problem for future work.
\section*{Acknowledgment}

 The first author would like to thank the University Grant Commission (UGC) (Ref. 221610074281), Govt. of India. The second author is thankful to the DST-INSPIRE (Ref. DST/INSPIRE/03/2023/000220), and the third author is thankful to, NBHM \linebreak(Ref. 0203/22(16)/2022/R$\And$D-II/11187), Department of Atomic Energy for their financial support. 
 \section*{Declaration}
 \subsection*{Author Contributions} All authors contributed equally to the writing of this paper.All authors read and approved
the final manuscript.
\subsection*{Conflict of interest}
The authors declare no conflict of interest.
\subsection*{Data availability}
The authors declare that all data supporting the findings of this study are available within the article.

\end{document}